\documentclass[reprint,showpacs,amsmath,amssymb,aps]{revtex4-1}

\usepackage{graphicx}
\usepackage{dcolumn}
\usepackage{bm}

\begin{document}

\title{Radiative Strength Functions for Dipole Transitions in $^{90}$Zr}

\author{I.D.~ Fedorets}
\email{fedorets@univer.kharkov.ua}
\author{S.S.~Ratkevich}%
\email{ratkevich@univer.kharkov.ua}
\affiliation{V.N.Karazin Kharkiv National University, Ukraine
}%


\begin{abstract}
Partial cross sections for the $(p,\gamma)$ reaction on the $^{89}$Y nucleus that
were measured previously at proton energies between 2.17 and 5.00 MeV and which
were averaged over resonances were used to determine the absolute values and
the energy distribution of the strength of dipole transitions from
compound-nucleus states to low-lying levels of the $^{90}$Zr nucleus.
The data obtained in this way were compared with the predictions of various models.
\end{abstract}


\maketitle

\section{\label{Intr}INTRODUCTION}

Resonance-like structures at excitation energies
well below the giant dipole resonance (GDR) were
discovered in the energy distributions of the $(E1,M1)$
dipole strength that were obtained from
total photoabsorption cross sections recently measured
for nuclei in the region around $A=90$ [1-6]. The
appearance of such structures is usually associated
with the fragmentation of the resonance or with the
excitation of other collective modes. Some of them
are observed predominantly in nuclei featuring closed
shells or nearly closed shells [1-3]. This is indicative
of the possible dependence of special features in the
energy distribution of the dipole strength on the nuclear structure.

From this point of view, it is of great interest to
study the distribution of the dipole strength in the
$^{90}$Zr nucleus. This rigid spherical nucleus featuring
a closed neutron shell $(N=50)$ is characterized by
a very high neutron separation energy of
$S_n=11970(3)$ keV, which is substantially higher
than the proton binding energy, $S_p=8354.8(16)$ keV
[7]. The total photoabsorption cross section studied
in [8] for $^{90}$Zr as a function of energy in
experiments with monochromatic photons in the energy
range between 8.5 and 12.5 MeV showed the presence
of a resonance-like structure that stands out
against the behavior obtained by extrapolating, to
this region, the giant dipole resonance approximated
by a standard Lorentz distribution. Specifically, an
obvious concentration of an additional strength was
observed around the energy of 11.5 MeV. The
experiments performed recently at new level of
sensitivity and accuracy by using bremsstrahlung
photons and reported in [2] revealed an excess of
the dipole strength around the energies of 6.5 and 9
MeV.

By studying the mechanism of formation of structures
in the same region of the energy distribution of the
dipole strength for the nucleus under study via
exciting this nucleus in different reactions, one
can ensure a more reliable identification of
observed structures and a more accurate quantitative
description of these structures. First of all, this
concerns reactions inverse to photoabsorption.
Within the traditional problem of a decrease in the
$E1$ strength over the low-energy wing of the giant
dipole resonance (especially below the
particle-emission threshold), it becomes necessary
in this case to study in more detail the
distribution of the strength of $E1$ transitions
between neutron-capture (proton capture) states and
levels of the final state nucleus at lower energies
with resort (for the sake of comparison) to data on
the photoexcitation of nuclei, which is a process
that relates the ground state to dipole and
quadrupole excitations.

Since a maximum of the $3p$-wave neutron strength
function is observed in $A\approx90$ nuclei, there
are reasons to assume that, in the case where these
nuclei are excited in $(n,\gamma)$ reactions, strong
valent $E1$ transitions in them are determined by
particle-hole configurations that are loosely bound
to respective GDR's [9] and which
may be one of the reasons for the appearance of the
structures that were discovered. In the presence of
nonstatistical effects, features that are observed
in the energy distribution of reduced partial
radiative widths averaged over various sets of
resonances and over the sets of both correlated and
uncorrelated transitions in $(n,\gamma)$ reactions
indicate that, in the region below the neutron
binding energy, the fragmentation of the giant
dipole resonance may not be a purely stochastic
process [10]. In contrast to what we have for other
nuclei whose mass numbers are close to $A=90$,the
results presented in [2,8] for $^{90}$Zr cannot be
supplemented with data for the respective
$(n,\gamma)$ inverse photoabsorption reaction
because of the absence of a stable target nucleus.
Despite the different roles that neutrons and
protons must play in the formation of the energy
distribution of the dipole strength in $^{90}$Zr, it
therefore becomes necessary to study it with the aid
of the $(p,\gamma)$ reaction on $^{89}$Y.

The objective of the present study is to determine
the absolute values and the energy distribution of the
strength of dipole transitions in the $^{90}$Zr nucleus in
the region of a possible manifestation of resonance-like
structures on the basis of already available experimental
data on partial and total cross section for the
reaction of proton radiative capture [11-16].

Investigation of special features of the energy
distribution of the dipole strength of gamma-transitions
in the $^{90}$Zr nucleus is of interest
not only as a source of unique information about the
structure of nuclei featuring a nearly closed shell
or the $N=50$ closed shell and about the effect of
this structure on the processes of photon emission
and absorption by such nuclei. Isotopes in the
region around $A=90$ are in the vicinity of one of
the maxima of the fission-fragment yield, and one
employs yttrium, zirconium, niobium, and molybdenum
in reactor structural materials. The energy
distribution of the dipole strength of gamma-transitions
in nuclei as expressed in terms of the
radiative strength function (RSF), which is a
fundamental feature of a nucleus in what is
concerned with photon absorption and emission by
this nucleus, is an indispensable component of a
quantitative analysis of cross sections for
photonuclear reactions and radiative nucleon capture
reactions inverse to them that is performed within
statistical theory. It follows that prospects for
the simulation of $p$-processes of nucleosynthesis
in nuclear astrophysics and the development of new
nuclear technologies depend crucially on reliably
determining the radiative strength function in the
energy region below the threshold for neutron
separation from nuclei of mass number close to
$A=90$. In relation to [17], where only the region
of probable values of the strength of dipole
transitions in $^{90}$Zr could be determined, the
present investigations are more detailed. A vast and
diverse set of experimental data accumulated thus
far serves here as a reliable basis for determining
parameters used in the statistical model.

\section{\label{sec2}EXPERIMENTAL RESULTS AND THEIR ANALYSIS}

In the present study, the energy distribution of the
strength of dipole gamma-transitions in the
$^{90}$Zr nucleus is determined in the reaction
$^{89}$Y$(p,\gamma_f)^{90}$Zr from a comparison of
the results obtained by measuring, at
incident-proton energies in the range between 2.17
and 5.0 MeV [11-14], and thereupon averaging partial
cross sections for this reaction with the radiative
strength function dependent cross section calculated
for this reaction on the basis of Hauser-Feshbach
statistical theory. This method for determining
RSF, which is described in
more detail in [18], for example, is
model-dependent, but the statistical model as such,
if justifiably applied, is one of the most precise
models for a quantitative analysis of nuclear data.
Its parameters either are well known or can be
refined in the same experiment for example, on the
basis of an analysis of the inelastic-scattering
channel, where the radiative channel plays an
insignificant role.

At incident-proton energies in the range of $E_p <
6$ MeV, the partial cross sections for the
$(p,\gamma)$ reaction on $^{89}$Y are determined
primarily by processes involving the production of a
compound nucleus [19] in the state $\lambda$ and its
deexcitation via direct gamma-transitions to lower
lying states f of the final nucleus $^{90}$Zr. The
ground-state spin-parity of the $^{89}$Y target
nucleus is $J^\pi_0=1/2^-$. In this nucleus, the
last unpaired proton is in the $2p_{1/2}$ shell. The
cross section for the $(p,\gamma)$ reaction in
question is dominated by the contribution of the
channel where the proton orbital angular momentum is
$l_p=0$ and $J^\pi_\lambda=1^-$ and which is
accompanied by the $E1$ transitions to the ground
state, whose quantum numbers are $0^+_1$; the first
excited state, whose quantum numbers are $0^+_2$ and
whose energy is 1760.71 keV; and to the $2^+$ states
at 2186.27, 3308.8, and 3842.2 keV. For the
$^{90}$Zr nucleus, all other final states such that
the intensities of primary gamma-transitions to them
can be measured, are characterized by a higher spin
and can reached only via the capture of protons
having higher values of the orbital angular momentum
$l_p$ or via a higher multipole order. The threshold
for the reaction $^{89}$Y$(p,n)^{89}$Zr is
$E_p=3.65$ MeV, but the ground state of the
$^{89}$Zr nucleus has the spin-parity of $J^\pi
=9/2^+$, and the transition to this state for slow
neutrons is improbable. The energy of $E_p=4.24$ MeV
at which the first excited state, whose quantum
numbers are $J^\pi=1/2^-$, corresponds to the
effective threshold. The heat of the $(p,\gamma_0)$
reaction on the $^{89}$Y nucleus is 8.3548(16) MeV;
therefore, the compound nucleus formed is excited to
the energy of 12.5 MeV at incident-proton energies
not exceeding the effective threshold for the
opening of the competing neutron channel (through
the reaction $^{89}$Y$(p,n_1)^{89}$Zr at $E_p=4.24$
MeV). At this excitation energy, the attained
density of states even in this even-even spherical
nucleus of $^{90}$Zr, which is magic in the number
of neutrons, $N=50$, exceeds (see, for example,
[20]) $10^4$ MeV$^{-1}$, which must satisfy the
requirements of applicability of a statistical
description. On the basis of an analysis of
investigations of the $(p,n)$, $(p,p)$, $(p,p')$,
and $(p,\gamma)$ reactions on the $^{89}$Y nucleus
and on neighboring nuclei [11,12,19-24],
$(\gamma,\gamma)$ and $(\gamma,\gamma ')$) reactions
on the $^{90}$Zr nucleus [2,25], as well as the
$(n,\gamma)$ reaction on the isotopes neighboring
$^{90}$Zr isotope [5,6], one can conclude that the
distances between resonance levels having identical
spins and parities in the product compound nucleus
$^{90}$Zr obey the Wigner distribution, while the
protonic and radiative widths follow two independent
Porter-Thomas distributions. At the same time, the
protonic width must be substantially larger than the
mean radiative width. Thus, there are reasons to
assume that conditions necessary for the
applicability of the statistical description hold.

The energy distribution of the strength of primary
gamma-transitions accompanying the deexcitation of
states of the compound nucleus can be described,
irrespective of the way of its formation, in terms
of quantities averaged over the corresponding
interval of excitation energies, such as the density
$\rho=D^{-1}_\lambda$ of states whose spins and
parities lie in specific intervals and radiative
strength functions; that is, the mean reduced
probability for an $E1$ or an $M1$ transition of
mean width $\Gamma^{XL}_{\lambda f}$ and energy
$E_\gamma$ connecting the $\lambda$ and $f$ states
is
\begin{equation}\label{f1}
S^{XL} \left( {E_\gamma  } \right) = {{\overline \Gamma  _{\lambda f}^{XL} } \mathord{\left/
 {\vphantom {{\overline \Gamma  _{\lambda f}^{XL} } {\left( {E_\gamma ^3 A^{{2 \mathord{\left/
 {\vphantom {2 3}} \right.
 \kern-\nulldelimiterspace} 3}} D_\lambda  } \right)}}} \right.
 \kern-\nulldelimiterspace} {\left( {E_\gamma ^3 A^{{2 \mathord{\left/
 {\vphantom {2 3}} \right.
 \kern-\nulldelimiterspace} 3}} D_\lambda  } \right)}}
\end{equation}
where $D_\lambda$ is the mean spacing between the
$\lambda$ states having a specific spin-parity.

According to statistical theory, the cross section
for the process in which the capture of a proton
with energy $E_p$ is followed by the emission of a
photon having an energy $E_\gamma$ and corresponding
to a primary transition from the state $\lambda$ of
the compound nucleus that reached equilibrium to the
final state f can be represented in the form
\begin{equation}\label{f2}
\sigma _{p\gamma }  = \frac{{\pi \mathchar'26\mkern-10mu\lambda _p^2 }}{{2(2I + 1)}}\sum\limits_J {(2J + 1)} \sum\limits_{l_p j_p } {\frac{{T_{l_p j_p } T_{\gamma \lambda f} }}{{T_\lambda  }}}
\end{equation}
where $\mathchar'26\mkern-10mu\lambda _p$ is the
reduced wavelength of the incident proton, $I$ is
the target spin, $J$ is the compound nucleus spin,
$T_{l_p j_p }$  is the coefficient of transmission
(sticking) for protons in the entrance channel,
$T_{\gamma \lambda f}$  is the coefficient of
transmission for photons of energy $E_\gamma =
E_\lambda - E_f$ that correspond to primary
transitions from the group of $\lambda$ states to
the final state $f$,and $T_\lambda$ is the sum of
transmission coefficients correspond- ing to all
open channels of deexcitation of $\lambda$ states.
Summation in expression (2) is performed over all
open reaction channels and over compound-nucleus
states whose quantum numbers J and $\pi$ are allowed
by relevant selection rules. In the present study,
the calculations are performed with allowance for
corrections for possible fluctuations of the cross
section that are due to a small number of open
channels [26-28]. At low energies, this correction
may prove to be significant. The transmission
coefficient $T_{\gamma \lambda f}$, which is the
quantity obtained by averaging, over compound
nucleus resonances, the probability for a $\gamma$
transition of multipolarity $L$, can be expressed in
terms of the partial RSF
$S_{\lambda f} (E_\gamma)$ as
\begin{equation}\label{f3}
T_{\gamma \lambda f}  = 2\pi S_{\lambda f} \left( {E_\gamma  } \right)E_\gamma ^{2L + 1}
\end{equation}
The coefficient $T_\lambda$ then has the form
\begin{equation}\label{f4}
\begin{array}{l}
 T_\lambda   = \sum\limits_{l_{p'} j_{p'} } {T_{l_{p'} j_{p'} } }  + \sum\limits_{l_n j_n } {T_{l_n j_n } }  +  \\
 \begin{array}{*{20}c}
   {} & {}  \\
\end{array} + 2\pi \sum\limits_J {\int\limits_0^{E_\lambda  } {\rho _J \left( {E_\lambda   - E_\gamma  } \right)S_{\lambda f} \left( {E_\gamma  } \right)E_\gamma ^{2L + 1} dE_\gamma  } }  \\
 \end{array}
\end{equation}
where ${T_{l_{p'} j_{p'} }}$ are the transmission
coefficients for protons in the exit channel;
${T_{l_n j_n } }$ are the transmission coefficients
for the neutron channel; and $\rho
_J(E_\lambda - E_\gamma)$ is the density of levels
characterized by a spin $J$,a parity $\pi$, and an
excitation energy $E_f$.

By using relations (2)-(4) and experimental data on
partial cross sections for respective $(p,\gamma)$
reactions, one can determine absolute values of the
RSF versus the photon energy
and properties of states between which the $\gamma$
transition being considered occurs. Because of the
dominance of the dipole mode in the radiative decay
of compound nucleus states, the quantity $
S_{\lambda f} \left( {E_\gamma  } \right) =
S_{\lambda f}^{E1} \left( {E_\gamma  } \right) +
S_{\lambda f}^{M1} \left( {E_\gamma  } \right)$,
which is the sum of radiative strength functions for
$E1$ and $M1$ transitions, is extracted from
experimental data on cross sections. The reliability
of the values obtained in this way for radiative
strength functions depends not only on the errors in
the experimental data used but also on the
justifiability and accuracy of the transmission
coefficients and level densities used in Eqs.
(2)-(4).

Our present calculations of the transmission
coefficients for protons relied on the traditional
optical model and employed as inputs the parameters
of the optical potential (OP) that were obtained
phenomenologically within one of the global
systematics compiled recently [29]. For
incident-proton energies below the Coulomb barrier,
the use of OP parameters directly from the global
systematics would be poorly justified, irrespective
of whether they were obtained on the basis of purely
phenomenological considerations or were calculated
on the basis of a realistic nucleon-nucleon
interaction and specific ideas of nuclear matter
[30,31]. By investigating the absorptive properties
of spherical nuclei in this region [22,24], it was
shown that, in the mass-number region around
$A\approx90$, the absorptive potential at subbarrier
energies is substantially smaller than that which
would follow from the global systematics. It would
be natural to assume [24] that, in the region of
closure of the $N=50$ shell, the decrease in the
absorptive potential, as well as the decrease in the
diffuseness parameter, as will be seen below, is due
to shell effects.

In the present study, the optical-potential
parameters from the global systematics in [29] that
were used as inputs were changed in such a way as to
obtain the best fit to available data on the cross
sections for the $(p,n)$ reactions on $^{89}$Y and
$^{93}$Nb nuclei for incident-proton energies in the
range from the neutron threshold to 5.8 MeV [22,23].
For some of the OP parameters already obtained, a
very small modification was needed in order to reach
subsequently the best fit to the cross section
measured for the $(p,p)$ and $(p,n)$ reactions on Zr
and Mo isotopes at incident-proton energies in the
range of $E_p=2\div7$ MeV [32], as well as to the
cross sections measured for the $(p,p)$, $(p,p')$,
and $(p,\gamma)$ reactions on $^{90}$Zr nuclei at
the incident-proton energies in the range of
$E_p=1.9\div5.7$ MeV [24]. All of the parameters
ultimately obtained for the real part of the optical
potential, with the exception of a somewhat reduced
diffuseness parameter $(a_R=0.62-0.73 fm)$, were
nearly identical to their counterparts in the global
systematics from [29], but the parameters of its
imaginary part differed markedly from those
presented in [29]. For example, the imaginary part
of the absorptive potential in the energy region
under investigation complied most closely with the
simple energy dependence (in MeV units) $W_D=2.73
+0.70E_p$. This corresponds to the change in $W_D$
from 4 MeV at $E_p=2$ MeV to about 7 MeV at $E_p=6$
MeV. For $E_p\geq 6$ MeV, the absorptive potential
approaches fast its values from the global
systematics.

In order to calculate the level density in the
$^{90}$Zr nucleus at excitation energies in the
range being studied, we employed either the
semiempirical back-shifted Fermi gas model or
microscopic methods. In the first case, the
calculations were performed with the level-density
parameter of $a =8.95(41)$ MeV$^{-1}$ and the energy
shift of $\Delta=1.97(30)$ MeV from the global
systematics recently published in [33] and obtained
on the basis of a fit to experimental data on the
schemes of levels in nuclei at low excitation
energies and on the distances between $s$-wave
neutron resonances at the neutron binding energy.
The data on the discrete section of the spectrum of
levels in $^{90}$Zr were taken from the NUDAT BNL
database [7], which is based on current publications
concerning this nucleus. Identical level densities
were assumed for positive- and negative-parity
states of identical spin. This assumption was
recently confirmed by good agreement of the results
calculated for the level density on the basis of the
back-shifted Fermi gas model with experimental data
obtained from the reaction
$^{90}$Zr($^3$He,$t)^{90}$Nb for the density of levels
corresponding to the $1^+$ states in the energy
range between 5 and 10 MeV, as well as with
experimental densities of $2^+$ and $2^-$ levels in
the $^{90}$Zr nucleus from high-resolution
measurements of the inelastic scattering of protons
with energy $E_p=200$ MeV at forward angles and
electrons with energy $E_0\approx66$ MeV at backward
angles on $^{90}$Zr [34].

Calculated and experimental level densities in the
$^{90}$Zr nucleus are given in Fig.1 versus the
excitation energy. The closed triangles and open
circles in Fig.1$a$ represent the level densities
obtained in [35] from the reactions
$^{90}$Zr$(p,p')$ and $^{90}$Zr$(e,e')$ for the
$2^+$ and $2^-$ states, respectively.
\begin{figure}[pt]
\includegraphics*[width=2.15in,angle=0.]{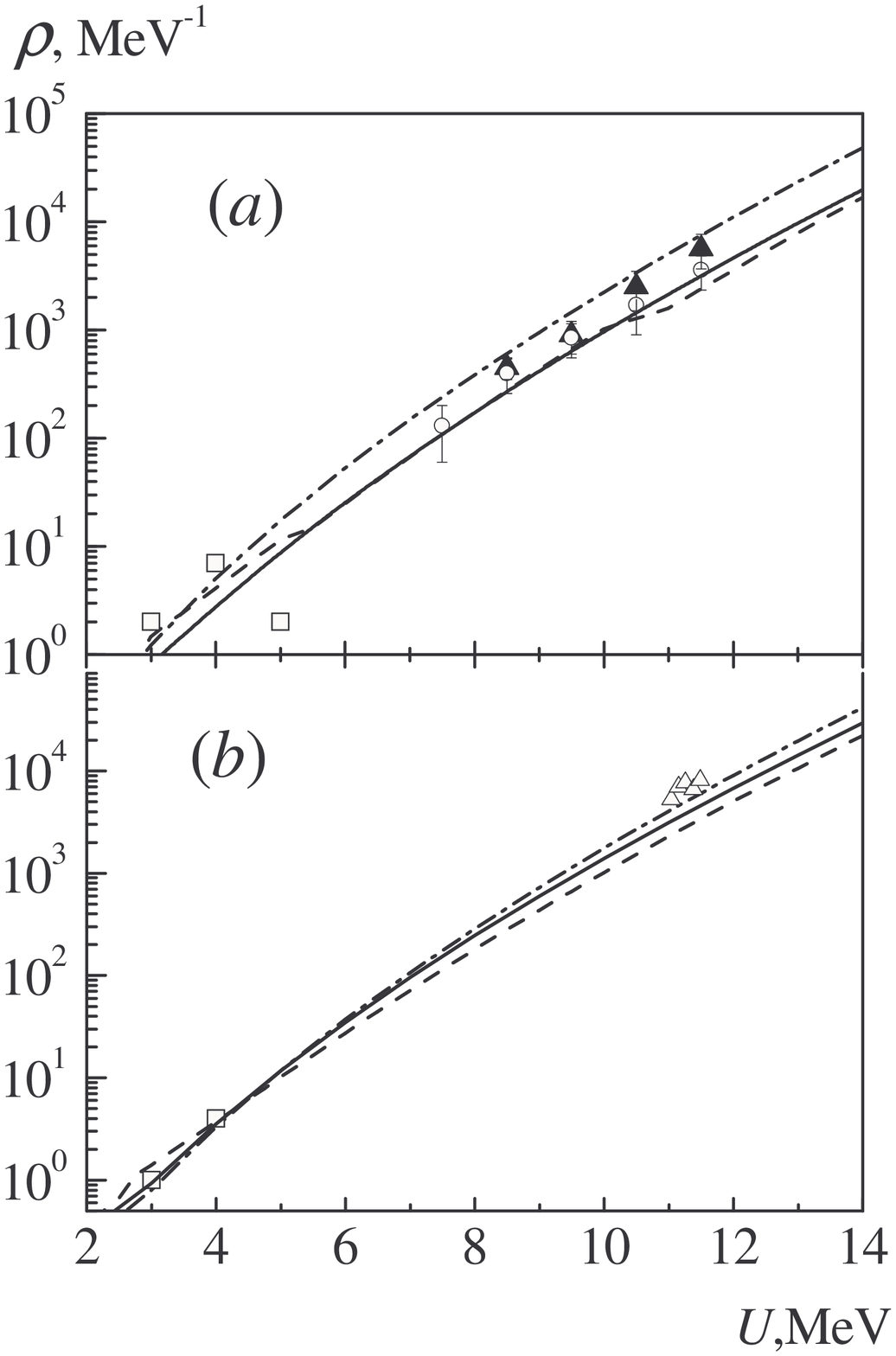}
\caption{\label{fig1}Level density in $^{90}$Zr as a function of the excitation energy for (\emph{a}) the $2^+$ and $2^-$
states and (\emph{b}) the $1^-$ states. The displayed points stand for (open boxes) experimental
data on discrete levels from [7]; (closed triangles and
open circles) data obtained for the $2^+$ and $2^-$ states, respectively, in [35] from the $(p,p')$ and $(e,e')$ reactions;
and (open triangles) data obtained for the $1^-$ states in [20] from the $(p,\gamma)$ reaction. The curves represent the level
densities calculated on the basis of the back-shifted Fermi gas model at (dash-dotted curve) the parameter values
of $\Delta=2.19$ MeV and $a =9.37$ MeV$^{-1}$ from [36] and (solid curve) the parameter values of $\Delta=1.97$ MeV and
$a =8.95$ MeV$^{-1}$ from [33] and (dashed curve) on the basis of the microscopic model from [38].}
\end{figure}
The open triangles
in Fig.1$b$ correspond to the density of levels for
the $1^-$ states that was obtained from a
statistical analysis of partial width fluctuations
manifesting themselves in the excitation function
measured for the reaction
$^{89}$Y$(p,\gamma_0)^{90}$Zr at incident-proton
energies in the range of $E_p=2.6\div3.3$ MeV [20].
The open boxes stand for data on discrete levels of
respective spin-parity. The solid curve in Fig.1
represents the results of the calculations based on
the back-shifted Fermi gas model and performed in
the present study at the parameter values of
$a=8.95$ MeV$^{-1}$ and $\Delta=1.97$ MeV. The
dash-dotted curve corresponds to the analogous
calculations, but the parameters were set there to
the values of $a=9.37$ MeV$^{-1}$ and $\Delta =
2.19$ MeV, which were obtained from an analysis of
the results of the parametrization used in [36] for
the level density in the region of $A \approx 90$
nuclei. The dashed curve stands for the results
available from [37], where the calculations relied
on the microscopic statistical model employing the
Hartreeâ-Fock + Bardeenâ-Cooper-Schrieffer (HF +
BCS) approximation [38].

The quantity $S_{\lambda f}(E_\lambda)$ [see Eq. (2)
and (3)] was chosen in such a way as to reproduce
the absolute experimental value of the partial cross
section for the $(p,\gamma)$ reaction. The radiative
strength function in expression (4) for the total
transmission coefficient was specified in a form
that was obtained in various theoretical approaches.
All of the remaining parameters obtained earlier
were fixed. It was assumed that the effect of the
model dependence on the values obtained in this way
for the partial RSF is weak
since, according to estimations, the mean radiative
widths are substantially smaller in this case than
the mean protonic widths. In order to determine the
absolute values of the RSF
for primary dipole transitions in the $^{90}$Zr
nucleus, we used the partial cross sections for the
$(p,\gamma)$ reaction that were measured by a
high-resolution method with thin $^{89}$Y targets in
the energy ranges of $E_p=2.2-3.4$ MeV [11] and
$E_p=2.17-5.0$ MeV [13,14] with a step close to the
equivalent target thickness. These cross sections
were averaged in such a way that the number of
compound nucleus resonances falling within the range
of averaging was sufficient for reducing to 5\% the
scatter of data that was due to Porter-Thomas
fluctuations. The isobaric analog states at 4.82 and
5.02 MeV, which were identified earlier, were
excluded from the averaging procedure. In Fig.2, the closed circles represent the values of the
RSF that were obtained from the averaged (over the interval of $\Delta E_p=1.2$
MeV) cross sections that were measured in [11] for the population of individual low-lying states of
$^{90}$Zr at incident-proton energies $E_p$ changing from 2.2 to 3.4 MeV with a step of 15 to 18 keV.
\begin{figure}[pt]
\includegraphics*[width=2.75in,angle=270.]{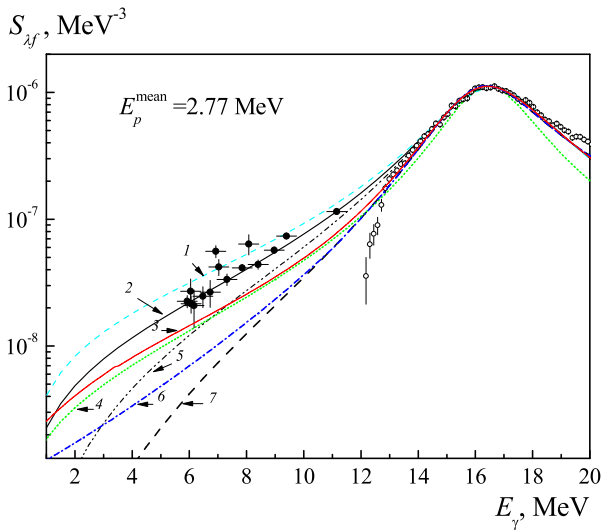}
\caption{\label{fig2}
Experimental and theoretical values of the radiative strength function for dipole gamma-transitions in $^{90}$Zr at the initial-
state energy fixed at $E_\lambda=11.1$ MeV and various values of the final-state energy $E_f$. The displayed points stand for (closed
circles) the RSF's obtained in our present study from the measured (in [11]) intensities of primary gamma-transitions in $^{90}$Zr
at $E^{mean}_p=2.77$ MeV for the averaging interval of width 1.2 MeV and (open circles) data on the giant dipole
resonance from [39]. The curves (for the respective notation, see main body of the text) correspond to \emph{1}-SLO, \emph{2}-MLO2,
\emph{3}-Sirotkin's approach, \emph{4}-GFL, \emph{5}-MLO1, \emph{6}-KMF, and \emph{7}-EGLO.}
\end{figure}
Averaging over so wide an interval was performed
because of strong fluctuations of original cross
sections in this energy range. In the measurements,
we used a $^{89}$Y target of equivalent thickness
13.7 keV at $E_p=3.0$ MeV. The energy range of
dipole transitions between 5.9 and 11.1 MeV for
which we extracted radiative strength functions was
determined by the range of energies of final states
for these transitions (from 0.0 to 5.2 MeV) at the
initial-state energy fixed at $E_\lambda=11.1$ MeV,
which corresponds to the mean incident-proton energy
of $E^{mean}_p=2.77$ MeV.

Employing the values obtained by measuring the
partial cross sections for the $(p,\gamma)$
reactions on $^{89}$Y nuclei at incident-proton
energies $E_p$ in the range between 2.17 and 5.0 MeV
[13,14] and thereupon averaging the results over the
range of $\Delta E_p =0.5$ MeV, we determined by the
same method the radiative strength functions for the
dipole transitions to the $0^+$ (0.0), $0^+$
(1.761), $2^+$ (2.186), $3^-$ (2.748), $2^+$ (3.309)
and $2^+$(3.842) states of the $^{90}$Zr nucleus
(the energies of the states in MeV units being
indicated parenthetically), the characteristics of
these states being established reliably.
\begin{figure}[pt]
\includegraphics*[width=2.75in,angle=270.]{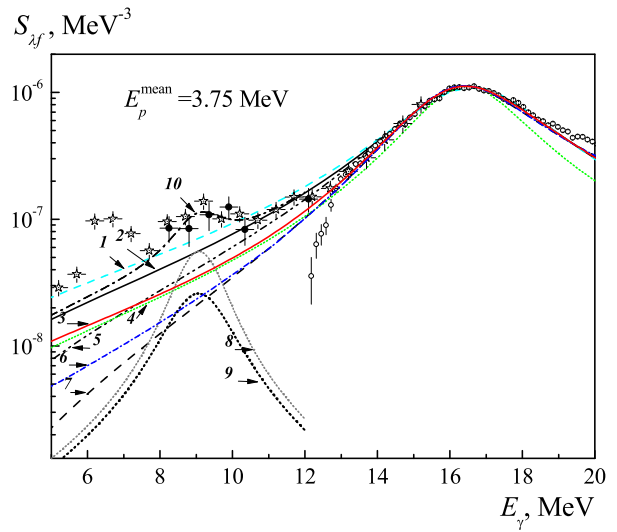}
\caption{\label{fig3}
As in Fig.2, but at $E_\lambda=12.1$ MeV. The displayed points stand for (closed circles) the RSF
values obtained in our present study from the measured (in [13, 14]) intensities of primary gamma-transitions in $^{90}$Zr
at $E^{mean}_p=3.75$ MeV, (stars) the RSF values from data on the the photoabsorption cross section from [2], and
(open circles) data on the GDR from [39]. Curves \emph{1-7} are identical to those in Fig.2; curves \emph{8} and \emph{9} represent
the hypothesized \emph{M}1- and \emph{E}1-resonance contributions, respectively, parametrized in terms of a Lorentzian distribution; and
curve \emph{10} corresponds to MLO2 supplemented with an \emph{M}1 resonance.}
\end{figure}
In Fig.3, the respective RSF values
obtained at the fixed energy of $E_\lambda=12.1$ MeV
($E^{mean}_p=3.75$ MeV) are represented by closed
circles. The open circles in Figs. 2 and 3
correspond to radiative strength functions extracted
the total cross sections presented in [39,40] for
the photoabsorption reaction in the giant dipole
resonance region, which receive a dominant
contribution from the $(\gamma,n)$ and $(\gamma,p)$
radiations. The stars in Fig.3 stand for the
RSF values obtained from the
total photoabsorption cross section measured in a
beam of bremsstrahlung gamma radiation [2]. Because
of strong fluctuations, those data were averaged
over the interval of $\Delta E_\gamma = 500$ keV.

\section{\label{Sec3}COMPARISON OF EXPERIMENTAL RESULTS WITH PREDICTIONS OF THEORETICAL MODELS}

From the assumption put forth in [8,41] that the
strength for both gamma-ray absorption and gamma-ray
emission in the region below the neutron separation
energy is determined by the low-energy tail of the
GDR, it follows that, at
incident-proton energies in the range being
considered, a virtual excitation of the giant dipole
resonance mode must have a decisive effect on the
exit $\gamma$ channel of the reaction
$^{89}$Y$(p,\gamma)^{90}$Zr. In this case, the
RSF for primary gamma-transitions of given multipolarity is related
through the detailed-balance principle to the
respective cross section for photoabsorption by a
nucleus in the ground state. It is obvious that, in
the region below the neutron binding energy in
$^{90}$Zr, the procedure of employing a unified
approximation of the RSF by
a Lorentzian distribution such that the energy of
its maximum coincides with the respective
experimental value is not justified theoretically,
since this procedure does not take fully into
account the observed structural effects.

The RSF values presented in Figs. 2 and 3 and
obtained from an analysis of experimental data are
contrasted against the radiative strength functions
calculated within various theoretical approaches
versus the photon energy decreasing to zero limit.
Curve 1 corresponds to the RSF related through the
detailed-balance principle to the photoabsorption
cross section in the form of a standard Lorentzian
distribution. The following parameters of the GDR
were used for the single-resonance Lorentzian
function: the energy at the maximum was $E_r=16.7$
MeV, the width was $\Gamma_r=4.2$ MeV, and the cross
section for photoabsorption at the maximum was
$\sigma_r = 211$ mb. These parameters were chosen in
such a way as to arrive at the best fit of a
Lorentzian distribution to experimental data
obtained in [39] for the energy dependence of the
photoabsorption cross section in the region of the
GDR maximum. Within the phenomenological standard
Lorentzian model (SLO in the notation adopted in
[37]), the absorptive width is taken to be a
constant that coincides with the GDR width. In the
calculations that are based on the use of a
generalized Lorentzian distribution (EGLO [37,42])
and whose results are represented by curve \emph{7}
in Fig.2 and in Fig.3, the absorptive width is
dependent on energy and is assumed to be
proportional to the collision component of the
damping width of zero sound in an infinite Fermi
liquid in the case where one takes into account only
a collision-induced two-body relaxation. Curves
\emph{2} and \emph{5} represent the radiative
strength functions calculated on the basis of the
modified Lorentzian models (MLO2 and MLO1,
respectively) [37,43], while curve \emph{4} stands
for the results of the calculations on the basis of
the generalized Fermi liquid model (GFL [44]).
Within these models, the absorptive width is also
dependent on energy, but, simultaneously, one takes
into account contributions both from fragmentation
and from collisions. These models of the radiative
strength function differ by the expression for the
absorptive width and by the contribution of various
dissipation mechanisms. Curve \emph{6} corresponds
to the results of the calculations for the radiative
strength function within the
Kadmensky-Markushev-Furman model [45] (KMF in the
notation adopted in [37]), which is based on taking
phenomenologically into account the coupling of
particle-hole configurations to more complex states.
Within this model, the expression representing the
spreading width of the GDR and corresponding to the
estimates obtained by Landau and Eliashberg for
damping within the theory of an infinite Fermi
liquid has the following form for even-even nuclei:
$ \Gamma \left( {E_\gamma  ,T_f } \right) = \Gamma
_r E_r^{ - 2} \left( {E_\gamma   + 4\pi ^2 T_f^2 }
\right) $, where $T_f$ is the effective temperature
of the nucleus in the state to which the
gamma-transition in question occurs, $\Gamma_r$ is
the width of the giant dipole resonance, and $E_r$
is the energy at its maximum. The results obtained
by calculating the radiative strength function
within Sirotkin's statistical approach [46], which
is based on Fermi liquid theory, are represented by
curve \emph{3}. A feature peculiar to the approach
described in [46] s that, in calculating the density
of $2p2h$-states, one takes into account the shell
structure of the spectrum of single-particle levels
and the effect of the nuclear temperature and
transition energy on the occupation probabilities
for these levels. Only states such that transitions
between them were allowed by the Pauli exclusion
principle were included in the calculation.

Figure 2 shows that the absolute values determined
for the RSF from
experimental data on partial cross sections for the
$(p,\gamma)$-reaction on $^{89}$Y at
$E^{mean}_p=2.77$ MeV do not deviate strongly from
an extrapolation of the Lorentzian distribution used
to parametrize experimental data in the region of
the GDR. Even at
$E^{mean}_p=3.75$ MeV (see Fig.3), however, the
radiative strength unctions obtained from the
$(p,\gamma)$ reaction in the energy region below 11
MeV show a sizable excess n absolute value above the
Lorentzian-like strength unction and a deviation in
shape from it. This agrees with the observation of
resonance-like structures above the extrapolation of
the Lorentzian distribution n the energy range
between 6 and 11 MeV in the photoabsorption reaction
[2,8]. Within a microscopic description of cross
sections for the photoabsorption of dipole gamma
rays by near-magic nuclei in the energy range of
$6\div8$ MeV, there are always remnants of intrinsic
transitions that existed in the model of
noncolliding particles. The authors of [2]
associated the concentration of strength in the
region of 6 to 7 MeV with the excitation of an
electric pygmy dipole resonance (PDR), but, at
higher energies, it becomes important to take into
account the coupling of pygmy and giant dipole
resonances to multiphonon states; it is difficult to
implement this in theoretical calculations, so that
one has to invoke various parametrizations. This
excess of strength, fragmented to a considerable
extent, can be exhausted, albeit not completely, by
a contribution hypothesized as in [5] and
parametrized in the form of a Lorentzian
distribution from the $M1$ resonance having an
energy of 9 MeV, a FWHM value of 2.5 MeV (1.2 MeV in
our present study), and a cross section of 7 mb at
the maximum. This contribution is represented by
curve \emph{8} in Fig.3. To a not lower degree of
plausibility, this could be an $E1$ resonance having
approximately the same parameters (curve \emph{9} in
Fig.3) and resulting from the possible
fragmentation of the GDR. The
results of experiments in beams of highly polarized
tagged photons [25] revealed that the strength that
can be associated with a giant $M1$ resonance in
$^{90}$Zr is broadly distributed over the
excitation-energy range from 8.1 to 10.5 MeV, which
is studied here. At the same time, a dominant
contribution of the $E1$-strength was identified in
this region.

We cannot rule out the possibility that a more
pronounced character of substructures observed in
the energy distribution of the radiative strength
function, which were estimated on the basis of
experimental data on the partial cross sections for
the $(p,\gamma)$-reaction on $^{89}$Y at
$E^{mean}_p=3.75$ MeV, is due not only to an
averaging interval narrower than that in Fig.2. To
a still greater extent, it may be caused by special
features present in original data. In Fig.4, the
experimental partial cross sections are contrasted
against their counterparts calculated by formula
(2).
\begin{figure}[pt]
\includegraphics*[width=2.75in,angle=0.]{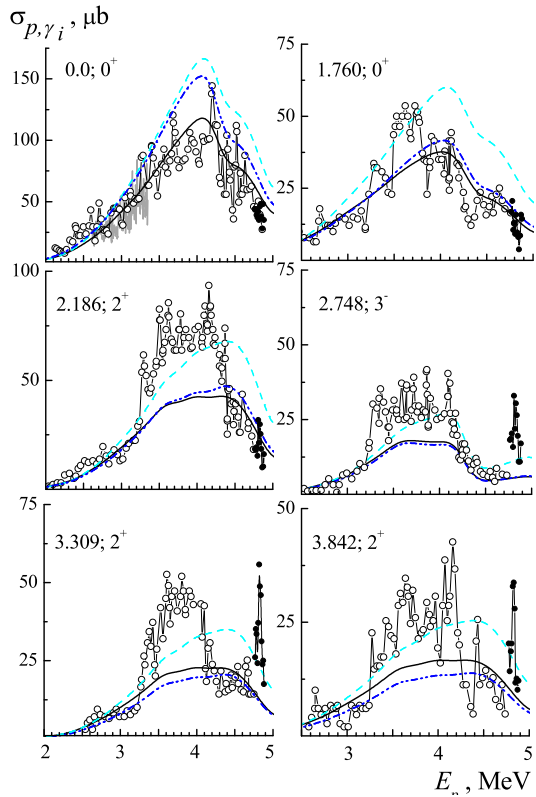}
\caption{\label{fig4}
Experimental partial cross sections for the $(p,\gamma_f)$ reaction for low-lying states of
$^{90}$Zr (the energies are given in MeV units) and their counterparts calculated within statistical theory. The open and closed circles stand for the experimental cross sections from[13] and [14], respectively. The curves represent the cross section calculated by using radiative strength functions found within the (dashed curve) SLO, (dash-and-double-dot curve) KMF, and (solid curve) MLO2 approaches.}
\end{figure}
In the calculations, use was made of radiative
strength functions obtained within various
theoretical approaches for $E1$ transitions to final
states of the $^{90}$Zr nucleus that were
characterized by specific spin-parities $J^\pi$. The
energies of states are given in MeV units. The open
circles represent the cross sections measured in
[13] at an angle of $56^\circ$ in the range of
$E_p=2.17\div4.80$ MeV. This corresponds to
excitation energies of the $^{90}$Zr nucleus in the
range between 10.5 and 13.1 MeV. The relative errors
in these measurements did not exceed a few percent.
The proton-energy step equivalent to the target
thickness changed from 20 to 25 keV. The closed
circles stand for the cross sections measured in
[14] at an angle of $55^\circ$ for $E_p$ values in
the range between 4.75 and 5.0 MeV with a step of 10
keV for a target of identical equivalent thickness.
A strongly fluctuating curve represents the cross
sections for the $(p,\gamma_0)$-reaction that were
determined after the respective normalization from
the relative intensities measured in [20] under the
aforementioned conditions for direct
gamma-transitions populating the ground state of the
$^{90}$Zr nucleus. The dashed, dash-and-double-dot,
and solid curves in Fig.4 represent the results
obtained by calculating partial cross sections with
the radiative strength functions obtained on the
basis of, respectively, the SLO, KMF, and MLO2
models. The results of a comparison with the
predictions of other models, which provide a poorer
description of experimental data on partial
radiative strength functions, are not presented in
Fig.4.

The measured excitation functions show a pronounced
fine structure not disappearing as $E_p$ increases
from 2.17 to 5.0 MeV and as the level density in the
compound nucleus grows accordingly, by a factor of
about six. After the averaging of the excitation
functions, for example, over an interval of width
270 keV, they become rather smooth in the range of
$E_p < 3.3$ MeV inclusive, apart from small
deviations for individual states, so that one can
describe them successfully within statistical
theory. However, variations belonging to the
intermediate-structure type and having different
shapes and magnitude for all final states, including
states characterized by identical $J\pi$, manifest
themselves at higher energies.

In describing the spectrum of levels of the
$^{90}$Zr nucleus on the basis of the simple shell
model, one most frequently takes 88Sr as an inert
core. At low excitation energies, the valent protons
in $^{90}$Zr must then be active in the $2p_{1/2}$
and $1g_{9/2}$ space and, to a smaller extent, in
the $1f_{5/2}$ and $2p_{3/2}$ space. The energy
difference between the $2p_{1/2}$ and $1g_{9/2}$
orbitals forms the energy gap for protons, which is
equal to 2670(90) keV [47], the pairing energy being
3593(8) keV. The shell-model gap for neutrons is
4445(8) keV, while the pairing energy amounts to
4093(12) keV. A strong mixing of the $\pi2p_{1/2}$
(up to 59\%)and $\pi1g_{9/2}$ (up to 41\%)
configurations both in the the ground state, whose
spin-parity is $0^+_1$, and in the first excited
state, whose spin-parity is $^0+_2$, is assumed in
this model. One also assumes that the
$\pi(1g_{9/2})^2$ configuration characterized by a
specific degree of mixing with excited-core states
corresponds to $J^\pi=2^+$ states. Possibly, the
$2^+_2$ and $2^+_3$ states feature a sizable
admixture of the $2p_{3/2}$ and $1f_{5/2}$
configurations [48]. The $3^-$ state is described as
an octupole phonon.

The fact that, at the proton energies in the region
of $E_p > 3.3$ MeV, the energy dependence of the
partial cross sections for the gamma-transitions to
all final states of the $^{90}$Zr nucleus in Fig.4,
with the exception of the $0^+_1$ and $0^+_2$
states, is not described within statistical theory
may be indicative, first of all, of the possible
contribution to the cross section for the
$(p,\gamma)$-reaction from the direct process of
single-particle transfer. This contribution as
manifested in cross sections for $(n,\gamma)$
reactions on $A\approx90$ nuclei is usually treated
as the result of an implementation of the valent
capture mechanism and the effect of doorway states
[49]. In this case, there are grounds to assume that
strong valent $E1$ transitions may be determined by
particle-hole configurations weakly coupled to the
corresponding GDR [7]. At the same time, the proton
strength function for $^{89}$Y in the energy range
under study is rather smooth [23,24], since the
single-particle $3s$ and $3p$ shape resonances are,
respectively, below 2 MeV and above 5 MeV. Moreover,
data on $(d,n)$ and $(^3$He,$d)$-reactions of
single-particle transfer (see [7] and references
therein) are insufficient for determining, to a
rather high degree of precision, the spectroscopic
factors for states of the $^{90}$Zr nucleus that are
of interest, but these spectroscopic factors are
necessary for a correlation analysis. In this
situation, it is difficult to explain in terms of a
mechanism acceptable for the $(n,\gamma)$ reaction
how that part of the radiative strength which is
associated with the simple component of the state
wave function may undergo fragmentation over
resonances (with allowance for their protonic
width). Nevertheless, special features of the
structure of the $^{90}$Zr nucleus do not rule out
the possibility for explaining the observed (in
Fig.4) nonstatistical contribution to the partial
cross sections for the $(p,\gamma)$ reaction on
$^{89}$Y in terms of the formation of doorway
states. We emphasize that, in the case of a dominant
valent contribution, the use of the intensities of
primary gamma-transitions from the reaction of
radiative neutron (proton) capture in order to
obtain data on radiative strength functions by the
method applied in the present study becomes
illegitimate.

On the other hand, the absence of nonstatistical
features in the partial cross sections for the
$(p,\gamma)$-reaction on $^{89}$Y could suggest
that, in this reaction at energies in the range
under study, there does not occur the excitation of
particle-hole states weakly coupled to the giant
dipole resonance, but the assumption of their
existence is crucial in explaining resonance-like
structures observed in the energy dependence of the
photoabsorption cross section, which correspond to a
manifestation of the mechanism of a predominantly
nonstatistical fragmentation of the giant dipole
resonance. In this connection, the problem of
comparing the strength of dipole transitions between
states populated via proton or neutron capture and
low-energy final-state levels corresponding to data
from experiments that studied nuclear
photoexcitation connecting the ground state to
dipole (quadrupole) excitations, as well as the
problem of performing a detailed comparison of the
mechanisms of $(p,\gamma)$- and
$(n,\gamma)$-reactions under conditions where
single-particle effects are operative, becomes
especially acute.

In order to analyze the total $(p,\gamma)$ cross
sections, which are determined by the total
radiative widths, it is necessary to know reliably
absolute values of radiative strength functions over
a broad energy range. The correctness of the choice
of model for RSF is
ultimately tested by addressing the question of
whether it can faithfully reproduce experimental
data over the entire range of gamma-transition
energies. The results obtained by calculating the
total cross sections for the $(p,\gamma)$-reaction
on $^{89}$Y nuclei on the basis of the
Hauser-Feshbach theory with radiative strength
functions found within various theoretical
approaches are given in Fig.5 along with
experimental data from [15,16].
\begin{figure}[pt]
\includegraphics*[width=2.15in,angle=0.]{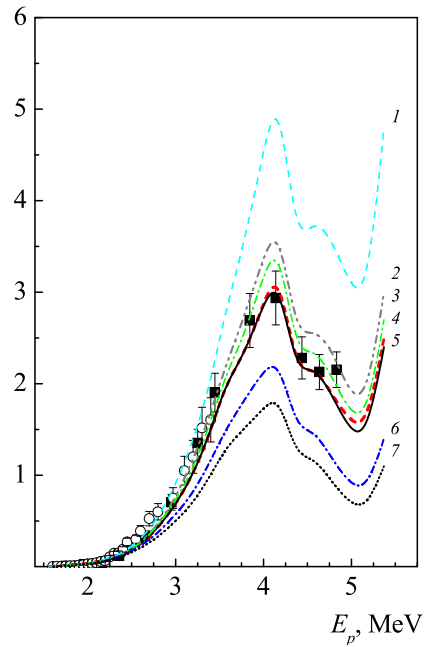}
\caption{\label{fig5} Total cross section for the
$(p,\gamma)$-reactionon $^{89}$Y. The open circles
and closed boxes stand for the experimental cross
sections from [15] and [16], respectively. The
curves represent the cross sections calculated on
the basis of Hauser-Feshbach theory by using the
radiative width found within various theoretical
models: \emph{1} - SLO, \emph{2} - microscopic model
from [50], \emph{3} - GFL, \emph{4} - Sirotkin's
approach, \emph{5} - MLO2, \emph{6} - KMF, and
\emph{7} - EGLO.}
\end{figure}
Curves \emph{1}-\emph{7} represent the results of these
calculations with the radiative strength functions
obtained on the basis of, respectively, the SLO
model, the microscopic model from [50], the GFL
model [44], Sirotkin's approach [46], the MLO2 model
[43], the KFM model [45], and the EGLO model [42].
From Fig.5, one can see that curve \emph{5}, which
was calculated with the RSF
based on the MLO2 model, and curve \emph{4}, which
is nearly coincident with it and which was
calculated with the RSF
based on Sirotkin's approach, exhibit the best
agreement with experimental data. At the same time,
the calculations with the radiative strength
function obtained on the basis of the standard
Lorentzian model yield total cross sections for the
$(p,\gamma)$-reaction on $^{89}$Y that are markedly
overestimated in relation to experimental data.

\section{\label{Sec4}CONCLUSIONS}

By using resonance-averaged data that were obtained
earlier from the $(p,\gamma)$-reaction on $^{89}$Y
[11-14], we have determined the absolute values and
energy distribution of the strength of dipole
transitions in the $^{90}$Zr nucleus in the region
of a possible manifestation of resonance-like
structures observed in the photoexcitation of this
nucleus [2,8]. This determination relies, first of
all, on the dependence of the cross section written
within statistical theory for the
$(p,\gamma)$-reaction on the spectroscopic
properties of the final- state nucleus and on its
statistical properties related to the radiative
strength function. The data obtained in this way are
compared with the results obtained by calculating
radiative strength functions within various
theoretical approaches.

The absolute values determined at the mean proton
energy of $E^{mean}_p=2.77$ MeV for the strength of
dipole transitions in the range of their energies
between 5.9 and 11.1 MeV do not deviate strongly
from the extrapolation of the Lorentzian
distribution that is used to parametrize
experimental data in the GDR
region. However, the RSF
corresponding to $E^{mean}_p=3.75$ MeV features an
obvious excess of the strength in agreement with the
resonance-like structure observed earlier in
experiments that studied photon scattering. We
cannot rule out the possibility that a more
pronounced character of the substructures observed
in this case is due to special features present in
original experimental data on partial cross
sections.

Additional measurements that would refine the level
of the nonstatistical contribution to the partial
cross sections for the $(p,\gamma)$-reaction on
$^{89}$Y nuclei and reliable experimental data on
spectroscopic factors from one-particle-transfer
reactions populating the same states of the
$^{90}$Zr nucleus as the respective
$(p,\gamma)$-reaction are required for drawing
definitive conclusions on the mechanism of formation
of the energy distribution of the strength of dipole
transitions in $^{90}$Zr.

The results of the calculations on the basis of
Hauser-Feshbach theory that employ the total
radiative width and which rely on the strength
functions obtained within the modified Lorentzian
model, Sirotkin's approach, the generalized Fermi
liquid model, and the microscopic model are in the
best agreement with experimental data on the total
cross sections for the $(p,\gamma)$-reaction on
$^{89}$Y nuclei. It is these models, but with
allowance for an excess strength parametrized in
terms of a Lorentzian distribution, that provide the
best description of experimental data on partial
strength functions as well. The use of a standard
Lorentzian distribution leads to total cross
sections substantially overestimated both in
relation to experimental data and in relation to the
calculations based on the different theoretical
models.

\end{document}